\documentclass[12pt]{article}
\usepackage{placeins}
\usepackage{amsmath,amsfonts,amssymb,bbm,natbib,amsthm}
\usepackage{fullpage}
\usepackage{calrsfs}
\usepackage{url}
\usepackage{bm}
\usepackage{booktabs}
\usepackage{graphicx}
\usepackage{relsize}
\usepackage{nicematrix}
\usepackage{eucal}
\usepackage{float}
\usepackage{subcaption}
\usepackage{threeparttable}
\usepackage{authblk}

\def\1{\mathbbm{1}}

\newtheorem{theorem}{Theorem}

\newtheorem{proposition}{Proposition}

\newcommand\HUGE{\@setfontsize\Huge{50}{60}}
\newcommand{\E}{\mathbb{E}}
\renewcommand{\P}{\mathbb{P}}
\newcommand{\R}{\mathbb{R}}

\newcommand{\dd}{\mathrm{d}}

\NiceMatrixOptions{cell-space-limits = 1pt}
\title{\LARGE Constructing Simultaneous Confidence Bands for Errors-in-variables Curves with Application to the Lorenz Curve} 
\date{}

\author[a]{Ziqing Dong}
\author[b]{Francesco Bartolucci}
\author[c]{Satoshi Kuriki}
\author[d,e]{Antonietta Mira \footnote{Corresponding Author}}
\affil[a]{Department of Statistics, University of California, Irvine, California 92697, United States (E-mail: ziqingd1@uci.edu)}
\affil[b]{Department of Economics, University of Perugia, Via A. Pascoli, 20, Perugia 06123, Italy (E-mail: francesco.bartolucci@unipg.it)}
\affil[c]{The Institute of Statistical Mathematics, 10-3 Midoricho, Tachikawa, Tokyo 190-8562, Japan (E-mail: kuriki@ism.ac.jp)}
\affil[d]{Faculty of Economics, Euler Institute, Università della Svizzera italiana, Via Giuseppe Buffi, 13, Lugano 6900, Switzerland (E-mail: antonietta.mira@usi.ch)}
\affil[e]{Department of Science and High Technology, University of Insubria, Via Valleggio, 11, Como 22100, Italy}

\begin{document}
\maketitle

\begin{abstract}
Errors-in-variables curves are curves where errors exist not only in the independent variable but also in the dependent variable.
We address the challenge of constructing simultaneous confidence bands (SCBs) for such curves. Our method finds application in the Lorenz curve, which represents the concentration of income or wealth.
Unlike ordinary regression curves, the Lorenz curve incorporates errors in its explanatory variable and requires a fundamentally different treatment.
To the best of our knowledge, the development of SCBs for such curves has not been explored in previous research. Using the Lorenz curve as a case study, this paper proposes a novel approach to address this challenge.
\end{abstract}

\noindent%
\noindent%

\newpage

\section{Introduction}
This paper develops a methodology of constructing simultaneous confidence bands (SCBs) for curves where errors are present in the explanatory variable, such as the Lorenz curve (\citeyear{lorenz1905methods}), the Bonferroni curve (\citeyear{bonferroni1930elementi}) and the ROC curve \citep{green1966signal}.
The Lorenz curve has attracted significant attention from researchers and has inspired numerous publications. \citet{kleiber2008lorenz} provides a review of the historical development and related research on the Lorenz curve, estimating that more than 500 methodological papers have been published on this topic in statistical and econometric journals.
Despite the extensive coverage of the topic in the literature, the construction of SCBs for the Lorenz curve, considering it as an errors-in-variables curve, appears to be an area that has not yet been explored.

Our interest originates from the limitations of the Gini index (\citeyear{gini1914sulla}) to measure income inequality, 
whose value is equivalent to two times the area between the line of perfect equality and the Lorenz curve. As highlighted by \citet{piketty2014capital}, summarizing the distribution of income using synthetic indices can be overly simplistic and misleading, as the significance of inequality varies across different parts of the distribution.

\begin{figure}[h]
  \centering
  \includegraphics[width=0.5\linewidth]{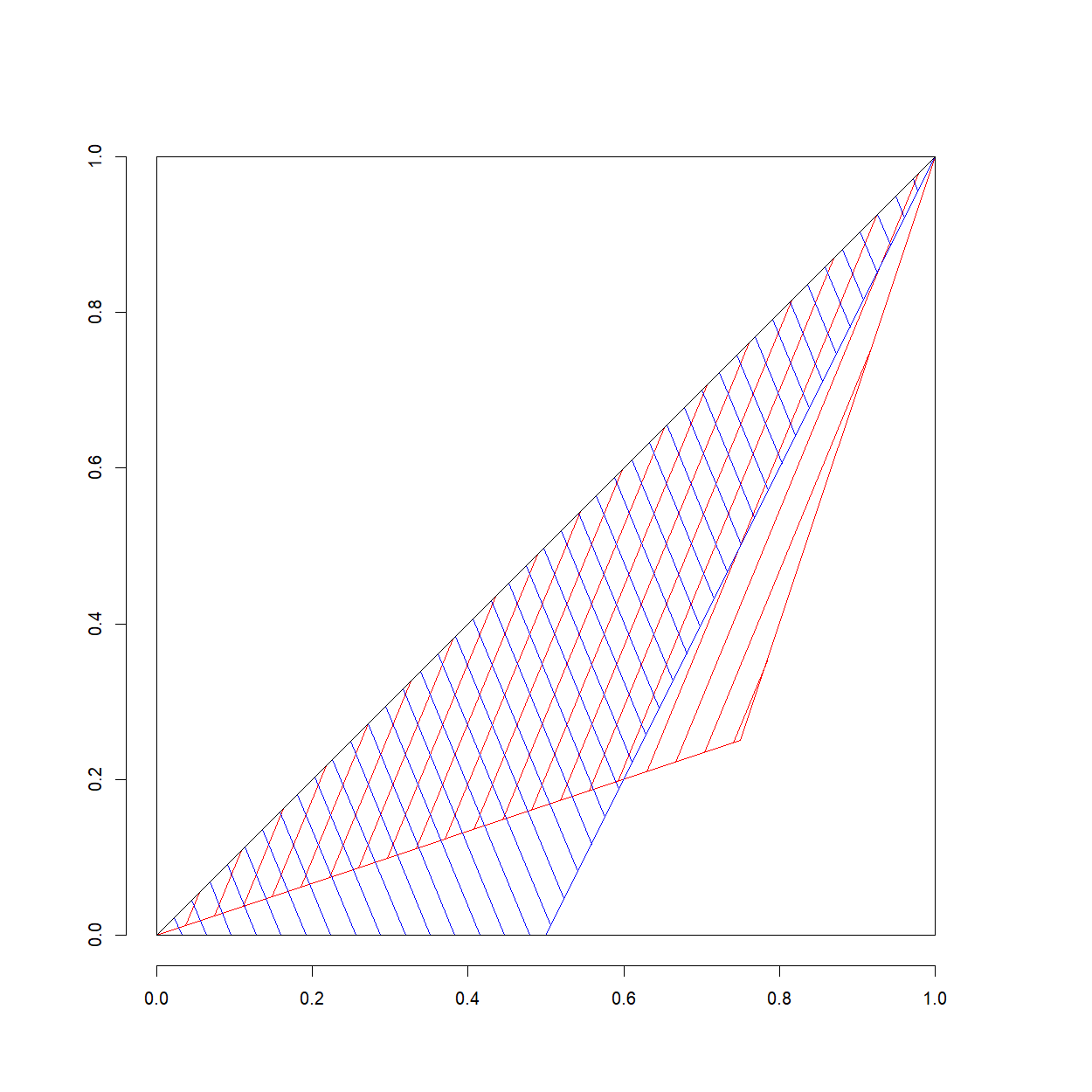}
\caption{The Gini indices calculated for the ‘‘red and blue economies’’ are both equal to twice the shaded areas of their respective representative colors, resulting in a value of 0.5 for both.}
  \label{gpair}
\end{figure}

The Lorenz curve provides more information than the Gini index, as illustrated by the following example.
Imagine two very different economies represented in terms of concentration in Figure~\ref{gpair} by the red and blue colors. 
Suppose that the ``red economy'' has three quarters of its population equally receiving one quarter of the total income while the rest one fourth of the population sharing the rest of the income, and the ``blue economy'' has half of its population having zero income and the rest half equally sharing the whole income. 
The Gini indices for both economies are equal to 0.5, and therefore, the Gini index fails to distinguish the inequality difference between the two economies.

Figure 1 provides a non-trivial example highlighting the limitations of the Gini index. This serves as one of the demonstrations of the need for graphical representations to complement synthetic statistics in order to better understand statistical problems. As stated by Lorenz in his original 1905 paper: ``the diagram will always tell what has happened" \citep{lorenz1905methods}. Indeed, by a visual inspection of the Lorenz curves in Figure~\ref{gpair}, one would hardly conclude that the income distribution of the ‘‘red and blue economies’’ are identical. Although the importance of statistical graphics is well understood, questions remain on how to estimate such curves on the basis of sample data and how to quantify the related uncertainty. 

In fact, in real-world applications, census data are difficult to obtain due to the financial, personnel and time constraints. Statistical surveys are often conducted from the samples to collect data for making inference on the targeted population. Coming back to the Lorenz curve example, the estimation of it from the sample data provides insights into the income distribution of the entire population. However, this process inevitably introduces errors.   

Given an estimate of a curve, one might naturally be interested in visualizing its estimation accuracy in all directions on a plane. The estimation of the Lorenz curve points causes uncertainty in both their abscissas and ordinates, and therefore, the estimated Lorenz curve is a curve which contains errors in its explanatory variable. The construction of SCBs for such curves requires a fundamentally different approach compared to ordinary regression curves \citep{working1929applications,scheffe1959analysis,liu2010simultaneous,delaigle2015confidence,kato2019uniform}, and remains unexplored in the existing literature. This paper proposes a generalization of the linearization method for obtaining the covariances and cross-covariances of the estimated Lorenz curve points. Since confidence ellipse is a generalization of confidence interval for quantifying the variability of a point on a two-dimensional plane, our idea of SCB is to make a union of the confidence ellipses of all the estimated points with individual confidence level adjusted for achieving the desired simultaneous confidence level. The adjustment of individual confidence level for building SCB is based on the result of \citet{davies1987hypothesis} for approximating the supremum of a chi-squared process.

The procedure first involves constructing point estimators for the curves, followed by deriving the covariances of these point estimators. Using the covariance information, we construct confidence ellipses for all point estimators based on a given sample. The SCB is formed by joining these confidence ellipses, with the individual confidence levels adjusted to ensure valid simultaneous inference.

This paper is organized as follows. In Section~\ref{sec:pro}, the 
general set-up and the Lorenz curve, along with its application and estimation, 
are presented. Section~\ref{sec:lin} introduces the Graf (\citeyear{graf2011use}) linearization method and estimates the variances of the Lorenz curve points, while a generalization of the linearization method is provided in Section~\ref{sec:cov} for estimating the covariances and cross-covariances of the Lorenz curve points. The theoretical construction of SCB for the Lorenz curve is presented in Section~\ref{sec:scb}. Section~\ref{sec:sim} studies the numerical applications of SCB under simple random sampling for different sample sizes and provides simulation results. Section~\ref{sec:con} concludes the paper.

\section{Preliminaries \label{sec:pro}}
\subsection{Set-up}
Many real-world studies involve finite populations, such as surveys, censuses, clinical trials, or agricultural studies. These populations are often of limited size and clearly defined (e.g., the population of a country, a set of schools, or a group of patients).
Let $N$ denote the size of the finite population of interest and $U=\{1,\ldots,N\}$ be the set of the unit labels.
Also let $y_1\leq y_2\leq \cdots \leq y_N$ be the ordered values of the variable of interest $y$, such as income, for every population unit.
The corresponding total, the 
partial total and the 
rank of $y$ 
are respectively
denoted by 
$Y=\sum_{i \in U} y_i$,
$Y_k=\sum_{i \in U} y_i \1{\scriptstyle[y_i \le y_k]}$ and 
$N_k=\sum_{i\in U}\1{\scriptstyle[y_i \le y_k]}$,
where $\1{\scriptstyle[\cdot]}$ is the indicator function.

For simplicity, only samples selected without replacement with a fixed sample size are considered. Consider a random sample $S$ of size $n$ ($n \leq N$) selected from $U$. 
A sampling design without replacement for the selection of the sample $s$, seen as a realization of $S$, with fixed sample size $n$ defines a probability mass function $p(\cdot)$ such that $\forall s \subset U, \, p(s)=\textrm{P}(S=s)$.

Let $a_1,\ldots,a_N$ 
be the Bernoulli random variables indicating the presence of the units in the random sample $S$, that is, $\forall i \in U$, $a_i=\1{\scriptstyle[i\in S]}$.  
By defining the sample indicator variables $a_i$'s, the values of the variable(s) of interest can be clearly separated from the source of the randomness of the $a_i$'s. The use of the sample indicator variables $a_i$'s was introduced by \citet{cornfield1944samples}, and is extensively used in this paper to estimate the variances and covariances of the points on the Lorenz curve.

Let $\pi_i$ be the first-order inclusion probability of unit $i$ and $\pi_{ij}$ be the second-order inclusion probability that units $i$ and $j$ are both selected in $S$. In statistical terms, we have that $\forall i \neq j \in U, \pi_i=\textrm{P}(i \in S)= 
\sum\limits_{\substack{s \subset U \\ i \in s}}p(s)=\E\left[a_i\right] \; \textrm{and}
\; \displaystyle \pi_{ij}=\textrm{P}(i \in S  \,\cap \, j \in S)=\sum\limits_{\substack{s \subset U \\ \{i,j\} \subset s}}p(s)=\E\left[a_ia_j\right].$
Specifically, consider the simple random sampling without replacement ($\textrm{SRSWOR}$) with fixed sample size: $\mathlarger{p(s)={N \choose n}^{-1}}.$ The first-order inclusion probabilities are:
$$\pi_i= 
\sum\limits_{\substack{s \subset U \\ i \in s}}p(s)=
{N-1 \choose n-1}{N \choose n}^{-1}=\frac{n}{N},  \forall i \in U,$$
and the second-order inclusion probabilities are:
$$\pi_{ij}= 
\sum\limits_{\substack{s \subset U \\ \{i,j\} \subset s}}p(s)=
{N-2 \choose n-2}{N \choose n}^{-1}=\frac{n(n-1)}{N(N-1)},  \forall i \neq j \in U.$$

A sampling weight $w_i$ may be 
associated to each population unit $i \in \! U$. Under the Horvitz-Thompson (\citeyear{horvitz1952generalization}) formulation, $w_i$ equals $1/\pi_i$ corresponding to the inverse of the first-order inclusion probability. The weights can also be determined by calibration. In principle, every $w_i$ can be interpreted as the number of units that unit $i$ represents in the population.

The plug-in estimators of the population size, the total, the partial total and the rank of $y$ are defined by
$\widehat{N}=\sum_{i \in U} w_i a_i$, $\widehat{Y}=\sum_{i \in U} w_i y_i a_i$, 
$\widehat{Y}_k=\sum_{i \in U} w_i y_i a_i \1{\scriptstyle[y_i \le y_k]}$ and
$ \widehat{N}_k= \sum_{i\in U} w_i a_i \1{\scriptstyle[y_i \le y_k]}$, respectively.

\subsection{The Lorenz curve\label{sec:curves}}
Given a non-negative random variable with cumulative distribution function $F(\cdot)$ and finite mean $\mu$, the Lorenz curve \citep{lorenz1905methods,gastwirth1971general} is theoretically defined as: 
$$
\mathcal{A}(p)=\mu^{-1}\int_0^p F^{-1}(t) dt, 0 \leq p \leq 1,
$$
where $F^{-1}(t)$ denotes the inverse of $F(t)$, defined as $F^{-1}(t)=\inf\limits_y\{y:F(y)\geq t\}$.  In real-life scenarios, the target population for inference is always finite. Given a measure of interest ($y$), the empirical Lorenz curve \citep{gastwirth1972estimation} is constructed through linear interpolation over $p \in [0,1]$ and is defined as:
$$
\mathcal{L}(p) = \frac{1}{\overline{Y}} \int_0^p F_N^{-1}(t) \, dt,  0 \leq p \leq 1,
$$
where $\overline{Y}$ represents the empirical mean and $F_N^{-1}$ denotes the inverse of the empirical distribution function.

Suppose $y$ represents income. In practice, $\mathcal{L}(p)$ ``plots along one axis cumulated per cents of the population from poorest to richest, and along the other the per cent of the total wealth held by these per cents of the population" as originally defined in Lorenz (\citeyear{lorenz1905methods}).
Let $\bm{L}_{i}= \left[L_{i(1)},L_{i(2)}\right]^\top$, where 
$L_{i(1)}= \frac{N_i}{N}$
and $L_{i(2)}= \frac{Y_i}{Y}$, $\forall i \in U$. 
$\bm{L}_i$ is the Cartesian coordinates of the $i^{th}$ point in a two-dimensional Euclidean space with $L_{i(1)}$ being the abscissa and $L_{i(2)}$ being the ordinate of $\bm{L}_i$. Let $\bm{L}_0=[0,0]^\top$ be the origin of a Euclidean space.
The Lorenz curve $\mathcal{L}(p)$ is equivalent to the piecewise linear interpolation of the $(N+1)$ points, $\bm{L}_0, \ldots, \bm{L}_N$,
within the unit square.

\begin{figure}[h]
  \centering
  \includegraphics[width=0.5\linewidth]{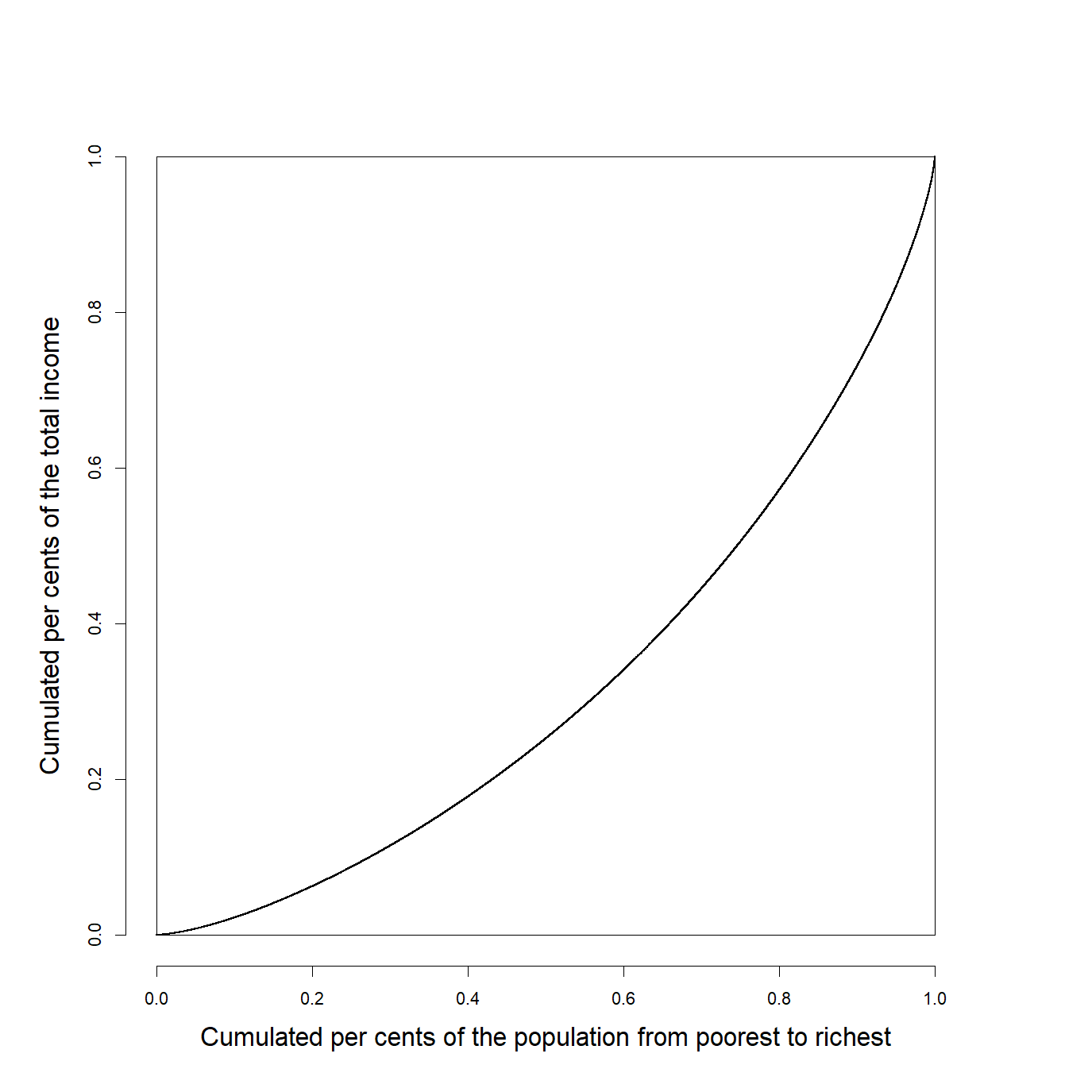}
\caption{The population Lorenz curve plotted using the 2015 IT-SILC data.}
\label{figlb}
\end{figure}

Figure~\ref{figlb} draws the Lorenz curve for the population using real data from the 2015 Italian component of the European Statistics on Income and Living Conditions (IT-SILC) \citep{istat2015indagini}. The 2015 IT-SILC is a survey on Italian household gross income in 2015. Because the 2015 IT-SILC dataset involves a very large number of units, we consider it as our population ($N$=17,942 households) and Figure~\ref{figlb} as the population Lorenz curve, which is hereafter the target of estimation.

When only the data referred to a sample of size $n$ are available, the population Lorenz curve will need to be estimated. 
The point estimators of the abscissas of the Lorenz curve following the plug-in principle are
\begin{equation*}
\widehat{L}_{k(1)}=\frac{\widehat{N}_k}{\widehat{N}}
=\frac{\sum\limits_{i \in U}w_ia_i\1{\scriptstyle[y_i \le y_k]}}{\sum\limits_{i \in U}w_ia_i}, \forall k \in S,
\end{equation*}
while the point estimators of the ordinates of the Lorenz curve of $L_{k(2)}$ are
\begin{equation*}
\widehat{L}_{k(2)}=\frac{\widehat{Y}_k}{\widehat{Y}}
=\frac{\sum\limits_{i \in U}w_iy_ia_i\1{\scriptstyle[y_i \leq y_k]}}{\sum\limits_{i \in U}w_iy_ia_i}, \forall k \in S.
\end{equation*}

A standard estimator of the Lorenz curve is the piecewise linear interpolation of the $n$ sample points plus the origin \citep{gastwirth1972estimation}. 

\section{Linearization and variance estimation of the Lorenz curve points \label{sec:lin}}
Estimating the variances and covariances of non-linear statistics, such as $\widehat{L}_{k(1)}$ and $\widehat{L}_{k(2)}$, is an intricate yet crucial step of building SCB for the Lorenz curve. Commonly, the variance of a non-linear statistic is estimated relying on the linearization methods \citep{lohr2021sampling}. 
In essence, linearization methods approximate the variance of a complex statistic by using the variance of the total estimator of the linearized variables. Generally speaking, various existing linearization methods give similar results. Among them, \cite{graf2011use} proposes linearization by differentiation of a statistic with respect to its sample indicator variables $a_k$'s, and is the simplest to apply \citep{tille2020sampling}. The Graf linearized variables depend on population parameters, which must therefore be estimated.

In the following, we introduce the linearized variables of the abscissas and ordinates of the point estimators for the Lorenz curve.
\begin{proposition}
\label{lin_Lk1}
The sample linearized variable of $\widehat{L}_{k(1)}$ is 
\begin{equation*}
\label{eq:lin_Lk1}
\hat{z}_{k(1	)i} :
=\frac{\partial \widehat{L}_{k(1)}}{\partial (w_ia_i)}
=\frac{\1\scriptstyle[y_i \leq y_k]}{\widehat{N}}
-\frac{1}{\widehat{N}}\widehat{L}_{k(1)}, \forall i \in U,
\end{equation*}
and its population counterpart is $z_{k(1)i}=\frac{\1\scriptstyle[y_i \leq y_k]}{N}-\frac{1}{N}L_{k(1)},\forall i \in U.$
\end{proposition}
\begin{proposition}
\label{lin_Lk2}
The sample linearized variable of $\widehat{L}_{k(2)}$ is 
\begin{equation*}
\label{eq:lin_Lk2}
\hat{z}_{k(2)i} :
=\frac{\partial \widehat{L}_{k(2)}}{\partial (w_ia_i)}
=\frac{y_i\1{\scriptstyle[y_i \leq y_k]}}{\widehat{Y}}
-\frac{y_i}{\widehat{Y}}\widehat{L}_{k(2)}, \forall i \in U,
\end{equation*}
and its population counterpart is
$z_{k(2)i} 
=\frac{y_i\1{\scriptstyle[y_i \leq y_k]}}{Y}
-\frac{y_i}{Y} L_{k(2)}, \forall i \in U.$  
\end{proposition}
\noindent
The proofs of Propositions~\ref{eq:lin_Lk1} and \ref{eq:lin_Lk2} are given in Appendix.

The linearized variables can be used for the variance estimation of each point estimator based on the Taylor expansion shown as follows:
\begin{align}
\nonumber \displaystyle\widehat{L}_{k(h)}\scriptstyle\left(w_1a_1,w_2a_2,\cdots,w_Na_N\right)
&\displaystyle\approx
\widehat{L}_{k(h)}\underbrace{\scriptstyle\left(1,\;1,\;\cdots,\;1\right)}_\textrm{N}
\displaystyle+\sum\limits_{i \in U}\frac{\partial \widehat{L}_{k(h)}\overbrace{\scriptstyle\left(1,\;1,\;\cdots,\;1\right)}^\textrm{N}}{\partial (w_ia_i)}\left(w_ia_i-1\right)\\ 
&=L_{k(h)}+\sum\limits_{i \in U}w_i z_{k(h)i}a_i-
\sum\limits_{i \in U}z_{k(h)i}, \forall h \in \{1,2\}. \label{eq:lin}
\end{align}
Rearrange Expression~\eqref{eq:lin}:
$\displaystyle \widehat{L}_{k(h)}-L_{k(h)} \approx \sum\limits_{i \in U}w_i z_{k(h)i}a_i-\sum\limits_{i \in U}z_{k(h)i}.$ Thus, $$\textrm{Var}\bigg[\widehat{L}_{k(h)}\bigg] \approx \textrm{Var}\bigg[\sum\limits_{i \in U}w_i z_{k(h)i}a_i\bigg].$$ By estimating $z_{k(h)i}$ by $\hat{z}_{k(h)i}$,
$\displaystyle \textrm{Var}\bigg[\widehat{L}_{k(h)}\bigg] \approx \textrm{Var}\bigg[\sum\limits_{i \in U}w_i \hat{z}_{k(h)i} a_i\bigg]$. 

Denote $\widehat{Z}_{k(h)} = \sum\limits_{i \in U}w_i \hat{z}_{k(h)i} a_i$. $\widehat{Z}_{k(h)}$ can be viewed as a total estimator.
For a general sampling design whose first-order ($\pi_i$) and second-order ($\pi_{ij}$) inclusion probabilities are all positive, a generalized Horvitz-Thompson formulation of the variance estimator of the total estimator $\widehat{Z}_{k(h)}$ is
\begin{equation}
\label{varht}
\widehat{\textrm{Var}}\bigg[\widehat{Z}_{k(h)}\bigg]=\sum_{i \in S} \sum_{j \in S} 
\frac{\hat{z}_{k(h)i}}{\pi_i} \frac{\hat{z}_{k(h)j}}{\pi_j} 
\frac{\pi_{ij} - \pi_i \pi_j}{\pi_{ij}}, \forall h \in \{1,2\}.
\end{equation} 
Simplification of Expression~\eqref{varht} under $\textrm{SRSWOR}$ results in the Horvitiz-Thompson variance estimator of the total estimator being:
\begin{equation}
\label{varsrs}
\widehat{\textrm{Var}}\bigg[\widehat{Z}_{k(h)}\bigg]=\frac{N^2}{n}\left(1-\frac{n}{N}\right)\frac{1}{n-1}\sum_{i \in S}\left(\hat{z}_{k(h)i}-\widehat{\overline{Z}}_{k(h)}\right)^2, \forall h \in \{1,2\},
\end{equation} where $\widehat{\overline{Z}}_{k(h)}$ is the mean of the sample linearized variables $\hat{z}_{k(h)i}$'s.

\section{Covariance matrix estimation of the Lorenz curve \label{sec:cov}}
In this section, the Graf linearization method is generalized to obtain the covariances of the Lorenz curve points based on the generalization of the Taylor's theorem.
For all $k \in S$, define  $\widehat{\bm{L}}_{k} = \left[\widehat{L}_{k(1)},\widehat{L}_{k(2)}\right]^\top$ as the point estimator of the Lorenz curve. $\widehat{\bm{L}}_{k}$ is thus a vector-valued function, $\widehat{\bm{L}}_{k} \colon \mathbb{R}^N \longrightarrow \mathbb{R}^2 $ , with the input  vector ${\displaystyle \mathbf{w \! \circ \! a}:\ =\left[w_1a_1,\ldots,w_Na_N\right]^\top},$ where $\mathbf{\circ}$ is the Hadamard product, ${\displaystyle\mathbf{w}:=[w_1,\ldots,w_N]^\top}$ is the vector composed of sampling weights and ${\displaystyle \mathbf{a}: =\left[a_1,\ldots,a_N\right]^\top}$ is the vector composed of the sample indicator variables. 
Let $\mathlarger{\mathbf{1}}$ be the $N$-dimensional column vector of 1's. 

Using the numerator layout, the derivative of $\widehat{\bm{L}}_{k}$ with respect to $\mathbf{w \! \circ \! a}$ is the $N \textrm{x} N$ Jacobian matrix $\mathlarger{\mathbf{J_k}}$ whose $(i,j)$th entry is the partial derivative, which by definition equals  $\hat{z}_{k(i)j}$:
$$\Bigr[\mathlarger{\mathbf{J}_k}\Bigr]_{ij}= \displaystyle\frac{\partial \widehat{L}_{k(i)}}{\partial (w_j a_j)}
=\hat{z}_{k(i)j}, \forall i \in \{1,2\}  \, \text{and} \; \forall j \in U.$$
The derivative of $\widehat{\bm{L}}_{k}$ at $\mathlarger{\bold 1}$ is represented by $\mathlarger{\mathbf{J}_k}\text{\small($\bold 1$)}$ whose $(i,j)$th entry, by definition, equals $z_{k(i)j}$:
$$\Bigr[\mathlarger{\mathbf{J}_k}\text{\small($\bold 1$)}  \displaystyle \Bigr]_{ij}= \displaystyle\frac{\partial \widehat{L}_{k(i)}}{\partial (w_j a_j)}\Bigg|_{\substack{\begin{subarray}{l}\mathlarger{\mathbf{w \hspace{-1mm} \circ \hspace{-1mm} a}}\\=
\mathlarger{\bold 1}\end{subarray}}}=z_{k(i)j},  \forall i \in \{1,2\}  \, \text{and} \; \forall j \in U.$$

The linearization of $\widehat{\bm{L}}_{k}$ relies on the generalization of Taylor's theorem :
\begin{align}
\label{eq:linv1} \displaystyle\widehat{\bm{L}}_{k}\mathbf{\left(w \! \circ \! a\right)}
&\displaystyle\approx
\widehat{\bm{L}}_{k}\text{\small($\bold 1$)} \displaystyle 
+\mathlarger{\mathbf{J}_k}\text{\small($\bold 1$)} \displaystyle \, \Bigl(\mathlarger{\mathbf{w}} \! \circ \! \mathlarger{\mathbf{a}} - \mathlarger{\bold 1} \Bigl)
\end{align}
Noticing $\widehat{\bm{L}}_{k}\text{\small($\bold 1$)} = \bm{L}_{k}$, rearrange Expression~\eqref{eq:linv1}:
$\displaystyle \widehat{\bm{L}}_{k}-\bm{L}_{k} \approx \mathlarger{\mathbf{J}_k}\text{\small($\bold 1$)}  \displaystyle \, \Bigl(\mathlarger{\mathbf{w}} \! \circ \! \mathlarger{\mathbf{a}} - \mathlarger{\bold 1} \Bigl).$ Thus, $$\textrm{Var}\biggr[\widehat{\bm{L}}_{k}\biggr] \approx \textrm{Var}\biggr[\mathlarger{\mathbf{J}_k}\text{\small($\bold 1$)} \displaystyle \, \Bigl(\mathlarger{\mathbf{w}} \! \circ \! \mathlarger{\mathbf{a}}\Bigl)\biggr]
=\mathlarger{\mathbf{J}_k}\text{\small($\bold 1$)}\displaystyle
\textrm{Var}\biggr[\Bigl(\mathlarger{\mathbf{w}} \! \circ \! \mathlarger{\mathbf{a}}\Bigl)\biggr] \mathlarger{\mathbf{J}_k^\top}\text{\small($\bold 1$)}.$$ 
Estimating $z_{k(i)j}$ by $\hat{z}_{k(i)j}$, $\forall i \in \{1,2\}  \, \text{and} \; \forall j \in U$,
$\displaystyle \textrm{Var}\biggr[\widehat{\bm{L}}_{k}\biggr] \approx \mathlarger{\mathbf{J}_k} \, \displaystyle
\textrm{Var}\biggr[\Bigl(\mathlarger{\mathbf{w}} \! \circ \! \mathlarger{\mathbf{a}}\Bigl)\biggr] \, \mathlarger{\mathbf{J}_k^\top}.$

Since
$\textrm{Var}\Big[\left(\mathlarger{\mathbf{w}} \! \circ \! \mathlarger{\mathbf{a}}\right)\Big]
=\E\Big[\left(\mathlarger{\mathbf{w}} \! \circ \! \mathlarger{\mathbf{a}}\right)\left(\mathlarger{\mathbf{w}} \! \circ \! \mathlarger{\mathbf{a}}\right)^\top\Big]
-\E\Big[\left(\mathlarger{\mathbf{w}} \! \circ \! \mathlarger{\mathbf{a}}\right)\Big]
\E\Big[\left(\mathlarger{\mathbf{w}} \! \circ \! \mathlarger{\mathbf{a}}\right)^\top \Big]$, 
$\textrm{Var}\Big[\left(\mathlarger{\mathbf{w}} \! \circ \! \mathlarger{\mathbf{a}}\right)\Big]$ can be represented by $\text{\large$\Lambda$}$ whose $(i,j)$th entry is
$$\Bigr[\text{\large$\Lambda$}\Bigr]_{ij}=\E\Bigr[w_iw_ja_ia_j\Bigr]-\E\Bigr[w_ia_i\Bigr]\E\Bigr[w_ja_j\Bigr], \forall i,j \in U.$$
The Horvitz-Thompson formulation under SRSWOR is: $\forall i,j \in U,$
\begin{equation*}
  \Bigr[\text{\large$\Lambda$}\Bigr]_{ij} =
    \begin{cases}
    \displaystyle \frac{N-n}{n}, & \text{if $i$ = $j$},\\[1em]
    \displaystyle -\frac{N-n}{n(N-1)}, & \text{if $i$ $\neq$ $j$}.
    \end{cases}       
\end{equation*}
$\textrm{{Var}}\Bigr[\mathlarger{\widehat{\bm{L}}_{k}}\Bigr]$ is the covariance matrix at the population level and needs to be estimated. We refer to Result 2.7 in \citet{tille2020sampling}[p. 21] for obtaining an unbiased estimator $\widehat{\textrm{{Var}}}\Bigr[\mathlarger{\widehat{\bm{L}}_{k}}\Bigr]$ of $\textrm{{Var}}\Bigr[\mathlarger{\widehat{\bm{L}}_{k}}\Bigr]$.

\begin{theorem}
The Horvitz-Thompson estimator of the covariance matrix $\textrm{{Var}}\Bigr[\mathlarger{\widehat{\bm{L}}_{k}}\Bigr]$ under SROSWOR is:
$$\widehat{\textrm{{Var}}}\Bigr[\mathlarger{\widehat{\bm{L}}_{k}}\Bigr]
=
\left[
\begin{NiceTabular}{l:l}
    $(\Sigma_{k,k})_{11}$ & 
    $(\Sigma_{k,k})_{12}$  \\ \hdottedline
    $(\Sigma_{k,k})_{21}$ & 
    $(\Sigma_{k,k})_{22}$
\end{NiceTabular}\right] 
=\mathlarger{\mathbf{\widehat{J}}_k} \, \displaystyle
\Delta \, \mathlarger{\mathbf{\widehat{J}}_k^\top},$$
where 
$$\Bigr[\mathlarger{\mathbf{\widehat{J}}_k}\Bigr]_{ij}=\hat{z}_{k(i)j}, \forall i \in \{1,2\} \; \mathrm{and} \; \forall j \in S,$$
and
$$
  \Bigr[\text{\large$\Delta$}\Bigr]_{ij} =
    \begin{cases}
    \displaystyle \frac{N-n}{n} \Big/ \pi_i =  \frac{N(N-n)}{n^2}, & \text{if $i$ = $j$} \in S,\\[1em]
    \displaystyle -\frac{N-n}{n(N-1)} \Big/ \pi_{ij} = -\frac{N(N-n)}{n^2(n-1)}, & \text{if $i$ $\neq$ $j$} \in S.
    \end{cases}    
$$
\label{th:cov}
\end{theorem}

The proof of Theorem~\ref{th:cov} is a simple consequence of the previous derivation and results. Further note that $(\Sigma_{k,k})_{11}$ and $(\Sigma_{k,k})_{22}$ are equivalent to $\widehat{\textrm{Var}}\bigg[\widehat{Z}_{k(1)}\bigg]$ and $\widehat{\textrm{Var}}\bigg[\widehat{Z}_{k(2)}\bigg]$ in Expression~(\ref{varsrs}) for all $k \in S$, which in part demonstrate Theorem~\ref{th:cov}.


The Lorenz curve points are not independent among each other. Using the generalized linearization method, the cross-covariances of the point estimators of the Lorenz curve can be derived. We estimate the cross-covariances of the neighbouring points, which are crucial for the construction of SCB.

Let $\mathlarger{\widehat{\bm{L}}}_{\frac{(k)}{(k+1)}}= \left[\widehat{\bm{L}}_{k},\widehat{\bm{L}}_{k+1}\right]^\top$ whose cross-covariance matrix is:
\begin{align*}
\text{Var}\Biggr[\mathlarger{\widehat{\bm{L}}}_{\frac{(k)}{(k+1)}}\Biggr]=\Bigg|
\begin{tabular}{ll}
    $\text{Cov}(\widehat{\bm{L}}_{k},\widehat{\bm{L}}_{k})$ & $\text{Cov}(\widehat{\bm{L}}_{k},\widehat{\bm{L}}_{k+1})$  \\ 
    $\text{Cov}(\widehat{\bm{L}}_{k+1},\widehat{\bm{L}}_{k})$ & $\text{Cov}(\widehat{\bm{L}}_{k+1},\widehat{\bm{L}}_{k+1})$  
\end{tabular}\Bigg|.
\end{align*}

The linearization of $\mathlarger{\widehat{\bm{L}}}_{\frac{k}{k+1}}$ is:
\begin{align}
\mathlarger{\widehat{\bm{L}}_{\frac{k}{k+1}}}\mathbf{\left(w \! \circ \! a\right)}
&\displaystyle\approx
\mathlarger{\widehat{\bm{L}}_{\frac{k}{k+1}}}\text{\small($\bold 1$)} \displaystyle 
+\mathlarger{\mathbf{J}_\frac{k}{k+1}}\text{\small($\bold 1$)} \displaystyle \, \Bigl(\mathlarger{\mathbf{w}} \! \circ \! \mathlarger{\mathbf{a}} - \mathlarger{\bold 1} \Bigl),
\end{align}
where $\mathlarger{\mathbf{J}_\frac{k}{k+1}}$ is a $4$x$N$ matrix represented as: $\mathlarger{\mathbf{J}_\frac{k}{k+1}}=\bigg[\displaystyle\frac{\mathbf{J}_k}{\mathbf{J}_{k+1}}\bigg]$.
$\forall i \in \{1,2\}  \, \text{and} \; \forall j \in U$, by estimating $z_{k(i)j}$ by $\hat{z}_{k(i)j}$ and $z_{k+1(i)j}$ by $\hat{z}_{k+1(i)j}$:
$$\textrm{{Var}}\biggr[\mathlarger{\widehat{\bm{L}}_{\frac{k}{k+1}}}\biggr] \approx \mathlarger{\mathbf{J}_{\frac{k}{k+1}}} \, \displaystyle
\textrm{Var}\biggr[\Bigl(\mathlarger{\mathbf{w}} \! \circ \! \mathlarger{\mathbf{a}}\Bigl)\biggr] \, \mathlarger{\mathbf{J}_{\frac{k}{k+1}}^\top}.$$

\begin{theorem}
Similar to Theorem~\ref{th:cov}, the Horvitz-Thompson formulation of the cross-covariance estimator $\widehat{\text{Var}}\Biggr[\mathlarger{\widehat{\bm{L}}}_{\frac{(k)}{(k+1)}}\Biggr]$ of $\text{Var}\Biggr[\mathlarger{\widehat{\bm{L}}}_{\frac{(k)}{(k+1)}}\Biggr]$ is: 
\begin{align*}
\widehat{\text{Var}}\Biggr[\mathlarger{\widehat{\bm{L}}}_{\frac{(k)}{(k+1)}}\Biggr]=\Bigg|
\begin{tabular}{ll}
    $\overbrace{\widehat{\text{Cov}}(\widehat{\bm{L}}_{k},\widehat{\bm{L}}_{k})}^{\Sigma_{k,k}}$ & $\overbrace{\widehat{\text{Cov}}(\widehat{\bm{L}}_{k},\widehat{\bm{L}}_{k+1})}^{\Sigma_{k,k+1}}$  \\ 
    $\underbrace{\widehat{\text{Cov}}(\widehat{\bm{L}}_{k+1},\widehat{\bm{L}}_{k})}_{\Sigma_{k+1,k}}$ & $\underbrace{\widehat{\text{Cov}}(\widehat{\bm{L}}_{k+1},\widehat{\bm{L}}_{k+1})}_{\Sigma_{k+1,k+1}}$  
\end{tabular}\Bigg|
=\mathlarger{\mathbf{\widehat{J}}_{\frac{k}{k+1}}^\top} \, \displaystyle
\Delta \, \mathlarger{\mathbf{\widehat{J}}_{\frac{k}{k+1}}^\top},
\end{align*}
where 
$\mathlarger{\mathbf{\widehat{J}}_\frac{k}{k+1}}=\bigg[\displaystyle\frac{\mathbf{\widehat{J}}_k}{\mathbf{\widehat{J}}_{k+1}}\bigg]$. In addition, $\Sigma_{k,k+1}=(\Sigma_{k,k+1})\displaystyle^{\top}=\Sigma_{k+1,k}$ and can be represented as: $\displaystyle \mathlarger{\mathbf{\widehat{J}}_k} \, 
\Delta \, \mathlarger{\mathbf{\widehat{J}}_{k+1}^\top}$.
\label{th:cross}
\end{theorem}
\noindent
The proof of Theorem~\ref{th:cross} is a simple consequence of the previous derivation and results.

\section{Simultaneous confidence band for the Lorenz curve \label{sec:scb}}
The coverage problem of the Lorenz curve around its endpoints is tricky and problem-specific. The point estimator of its left endpoint (i.e., $\bm{L}_0$) takes always the value  $(0,0)$ and is a degenerate random variable, which has no variance-covariance structure, while the point estimator of its right endpoint (i.e., $\bm{L}_N$) takes always the value $(1,1)$, which is usually biased and has an asymmetric distribution as the curve is bounded within the unit square. We hereafter focus on covering the main part of Lorenz curve (not around its endpoints) in order to develop a generic method of constructing SCB for errors-in-variables curve. 

Assume that $\{\widehat{\bm{L}}_k\}_{k=1,\ldots,(n-1)}$ are jointly distributed as a bivariate Gaussian distribution with mean $\bm{\mu}_k=\left[\mu_{k(1)},\mu_{k(2)}\right]^\top$ and covariance structure
$
\mathlarger{\Sigma}_{k,k},
$
where $\bm{\mu}_k$'s are unknown and $\Sigma_{k,k}$'s are known. The normality assumption can be justified on the basis of large sample theory arguments \citep{wilks1962mathematical,rao1973linear,beach1985joint}.
Let the points $\{\bm{\mu}_{k}\}_{k=1,\ldots,(n-1)}$ be the true Lorenz curve points and $\{\widehat{\bm{L}}_k\}_{k=1,\ldots,(n-1)}$ be their point estimators.
Suppose that the true Lorenz curve is the piecewise linear curve connecting neighbours:
\begin{equation}
 \mathcal{L}(t) = (1-\delta)\bm{\mu}_{k} + \delta\bm{\mu}_{k+1} \in\R^2, \quad t \in [1,(n-1)],
 \label{eq:lcurve}
\end{equation}
where
\[
 k=k(t)=[t], \quad \delta=\delta(t)=t-[t].
\] 
The estimator for $\mathcal{L}(t)$ is
\begin{equation}
 \widehat{\mathcal{L}}(t) = (1-\delta)\widehat{\bm{L}}_k + \delta \widehat{\bm{L}}_{k+1} \in\R^2, \quad t\in [1,(n-1)].
  \label{eq:lhatcurve}
\end{equation}
Then,
\[
 \widehat{\mathcal{L}}(t) \dot{\sim} N_2\big(\mathcal{L}(t) ,\Sigma(t)\big) \quad \text{and} \quad
 \Sigma(t) = \big((1-\delta)I_2 , \delta I_2\big)
 \begin{pmatrix} \Sigma_{k,k} & \Sigma_{k,k+1} \\ \Sigma_{k+1,k} & \Sigma_{k+1,k+1} \end{pmatrix} \begin{pmatrix} (1-\delta)I_2 \\ \delta I_2 \end{pmatrix}.
\]

Define a chi-square process
\[
 g(t) = \big(\widehat{\mathcal{L}}(t) -\mathcal{L}(t) \big)^\top\Sigma^{-1}(t)\big(\widehat{\mathcal{L}}(t) -\mathcal{L}(t) \big), \quad t\in [1,(n-1)].
\]
For each $t$, $g(t)\sim\chi^2_2$.
The tail probability formula for $\sup_{t\in [1,(n-1)]}g(t)$ is proposed
by Eq.\ (3.2) of \citet{davies1987hypothesis}.
Let $c_{\alpha}$ be the point such that
\[
 \P\left(\sup_{t\in [1,(n-1)]} g(t) < c_{\alpha}\right) =
 \P\Bigl(g(t) < c_{\alpha}, \forall t\in [1,(n-1)]\Bigr) = 1-\alpha.
\] 
Then, SCB for $\mathcal{L}(t) $ is obtained as
\begin{align}
\bigcup_{t\in [1,(n-1)]}
\left\{\mathcal{L}(t)  \mid \big(\widehat{\mathcal{L}}(t) -\mathcal{L}(t) \big)^\top\Sigma^{-1}(t)\big(\widehat{\mathcal{L}}(t) -\mathcal{L}(t) \big) < c_\alpha
 \right\}.
 \label{eq:expscb}
\end{align}
Meanwhile, the explicit form of the point-wise confidence band (PCB) for $\mathcal{L}(t) $ can be expressed as 
\begin{align}
\bigcup_{t\in [1,(n-1)]}
\left\{\mathcal{L}(t)  \mid \big(\widehat{\mathcal{L}}(t) -\mathcal{L}(t) \big)^\top\Sigma^{-1}(t)\big(\widehat{\mathcal{L}}(t) -\mathcal{L}(t) \big) < d_\alpha
 \right\}
 \label{eq:exppcb}
\end{align}
with $d_\alpha$ being the $(1-\alpha)$ quantile of $\chi^2_2$.

Recall that the assumed true curve $\mathcal{L}(t) $ and its estimator $\widehat{\mathcal{L}}(t) $ are given in (\ref{eq:lcurve}) and (\ref{eq:lhatcurve}), respectively.
The variance of $\widehat{\mathcal{L}}(t) $ is $\Sigma(t)$.
When $t$ is fixed,
\[
 g(t) = (\widehat{\mathcal{L}}(t) -\mathcal{L}(t) )^\top\Sigma^{-1}(t)(\widehat{\mathcal{L}}(t) -\mathcal{L}(t) )
\]
is distributed as the chi-square distribution with 2 degrees of freedom
\[
 \P\bigl(g(t)>u\bigr) = \P(\chi^2_s>u), \quad s=2. 
\]

From \citet{davies1987hypothesis} (see Theorem A.2),
\begin{align*}
 \P\biggl(\sup_{t\in[1,(n-1)]}g(t)>u\biggr)
 &\sim \P(\chi^2_s>u) + C \times u^{\frac{1}{2}(s-1)}e^{-\frac{1}{2}u}\pi^{-\frac{1}{2}}2^{-\frac{1}{2}s}/\Gamma(\frac{1}{2}s+\frac{1}{2}) , \quad s=2,
\end{align*}
where
\begin{align*}
 C = \int_1^{n-1}\E[\Vert\eta(t)\Vert]\dd t, \quad 
 \E[\Vert\eta(t)\Vert]
 = \E[|\partial g(t)^{\frac{1}{2}}/\partial t|]\pi^{\frac{1}{2}}\frac{\Gamma(\frac{1}{2}s+\frac{1}{2})}{\Gamma(\frac{1}{2}s)}.
\end{align*}
Estimation of $\E[|\partial g(t)^{\frac{1}{2}}/\partial t|]$ can be obtained by simulation. Detailed calculation of $\partial g(t)^{\frac{1}{2}}/\partial t$ is provided as follows:
\begin{align*}
\partial g(t)^{\frac{1}{2}}/\partial t = 2 \big(\dot{\widehat{\mathcal{L}}}(t) -\dot{\mathcal{L}}(t) \big)^\top \Sigma^{-1}(t)\big(\widehat{\mathcal{L}}(t)-\mathcal{L}(t) \big) + 
\big(\widehat{\mathcal{L}}(t) -\mathcal{L}(t) \big)^\top \dot{\Sigma}^{-1}(t)\big(\widehat{\mathcal{L}}(t) -\mathcal{L}(t) \big),
\end{align*}
where $\dot{\Sigma}^{-1}(t)=-\Sigma^{-1}(t)\dot{\Sigma}(t)\Sigma^{-1}(t).$

\section{Numerical implementation and simulation \label{sec:sim}}
In the present section, the numerical implementation and simulation results of SCB construction for the Lorenz curve are studied. As mentioned in Section~\ref{sec:curves}, we consider the real 2015 IT-SILC data as population to validate our method of building SCB presented in Section~\ref{sec:scb}. We artificially draw samples from the population using SRSWOR in order to test the performance of our method of SCB construction. The population Lorenz curve plotted in Section~\ref{sec:curves} is our target of estimation. 

As Lorenz curve should be bounded within the lines of perfect equality and inequality, SCB which goes beyond these two lines is cut numerically for practical purposes. This procedure does not affect the accuracy on the coverage of the constructed SCB, but adds aesthetics and practicality to the plot. 

Table~\ref{alg} presents the procedure of estimating the Lorenz curve and building SCB for the estimated Lorenz curve. The algorithm is written in a generic format for building SCBs of curves with errors-in-variables using the Lorenz curve as an example.

\begin{table}[htb!]
\centering
\begin{threeparttable}
\caption{Algorithm for constructing SCB for errors-in-variables curves —— using the Lorenz curve as an example}
\begin{tabular}{lrrrrrr}
\toprule
1: Select a sample according to a sampling design \\
2: Based on the selected sample, build point estimators $\widehat{L}_{k(1)}$ and $\widehat{L}_{k(2)}$, $\forall k \in S$\\
3: Interpolate the point estimators to obtain an estimator of the Lorenz curve  \\
4: Linearize the point estimators to estimate the variance matrix  $\Sigma (t)$,  $\forall t \in [1,(n-1)]$\\
5: Simulate C to approximate the empirical $(1-\alpha)$ quantile  $u_{\alpha}$ of $\sup_{t\in [1,(n-1)]}g(t)$ \\
6. Plot the unions of the confidence ellipses of $\widehat{\mathcal{L}}(t), \forall t \in [1,(n-1)]$ based on the adjusted\\ \;\;\;critical value $u_{\alpha}$ as SCB of $\widehat{\mathcal{L}}(t)$ at the confidence level $(1-\alpha)$ \\
\bottomrule
\end{tabular}
\label{alg}
\end{threeparttable}
\end{table}

\begin{figure}[!ht]
\centering
\begin{subfigure}{.45\textwidth}
  \centering
  \includegraphics[width=\linewidth]{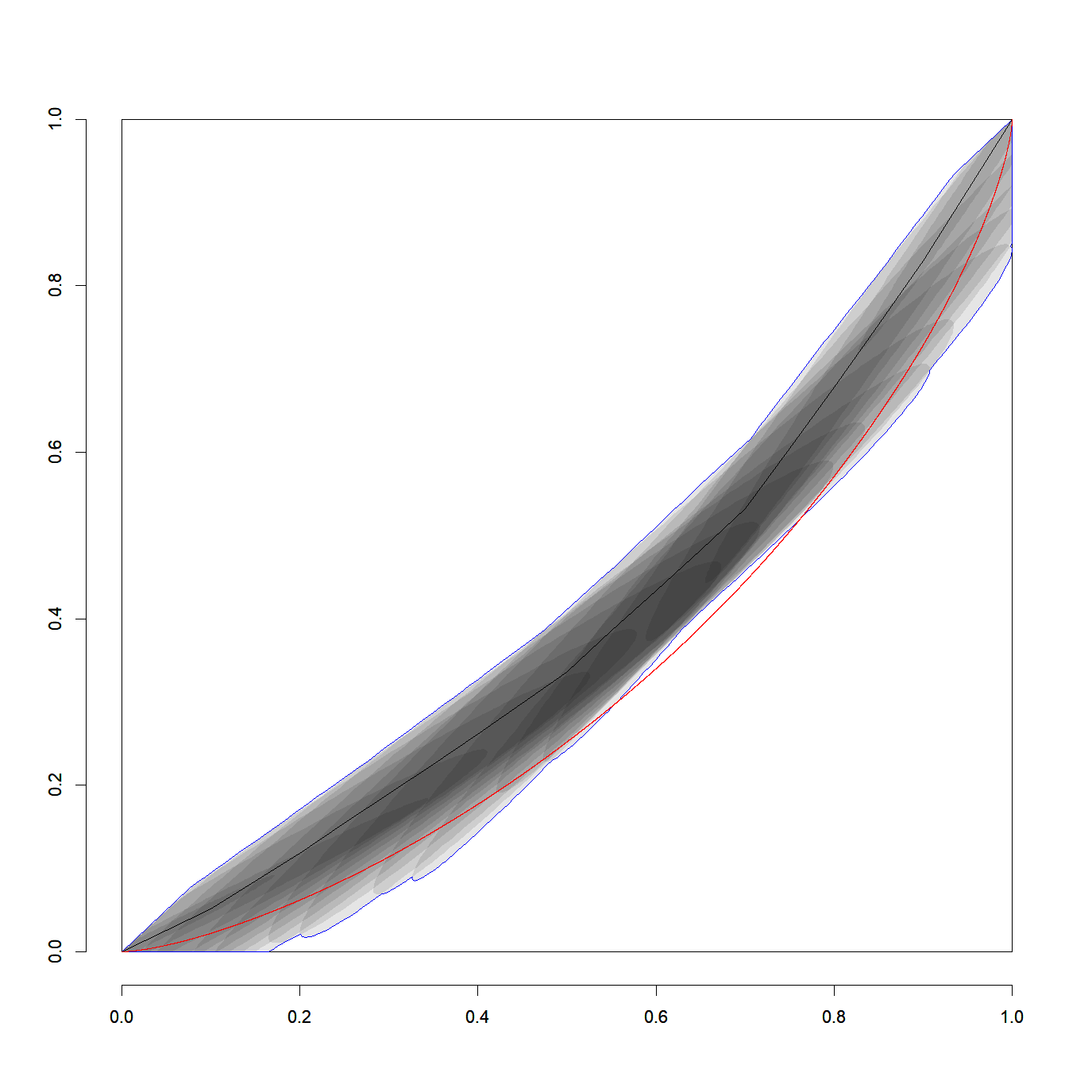}
  \caption{PCB (n=10, $c_\alpha=5.991$).}
  \label{fig:expcb}
\end{subfigure}%
\begin{subfigure}{.45\textwidth}
  \centering
  \includegraphics[width=\linewidth]{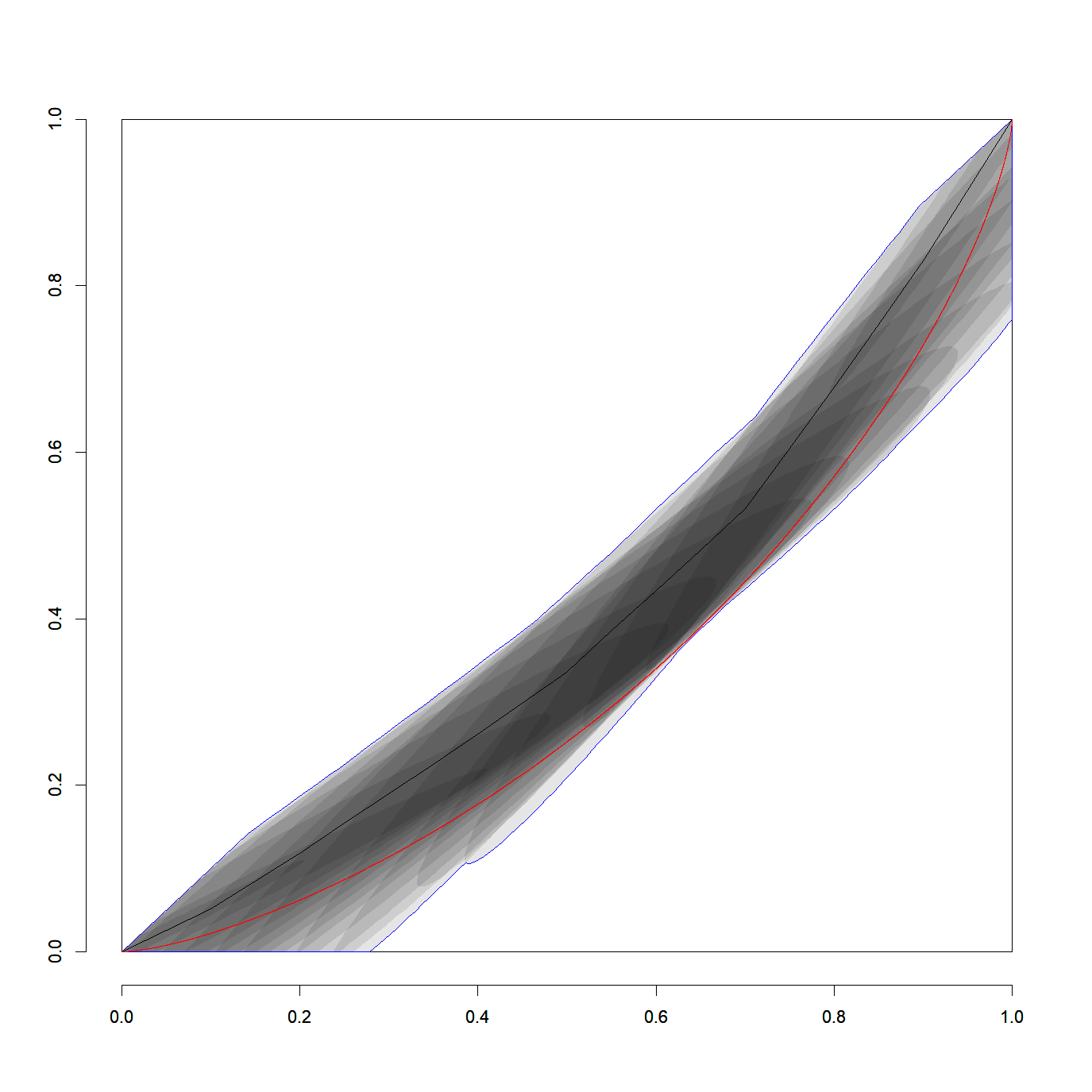}
  \caption{SCB (n=10, $u_\alpha=9.689$).}
  \label{fig:exscb}
\end{subfigure}
\caption{PCB and SCB constructed based on a sample of size 10 drawn by SRSWOR using the 2015 IT-SILC data at 95\% confidence level.}
\label{fig:10p}
\end{figure}

\begin{figure}[!ht]
\centering
\begin{subfigure}{.45\textwidth}
  \centering
  \includegraphics[width=\linewidth]{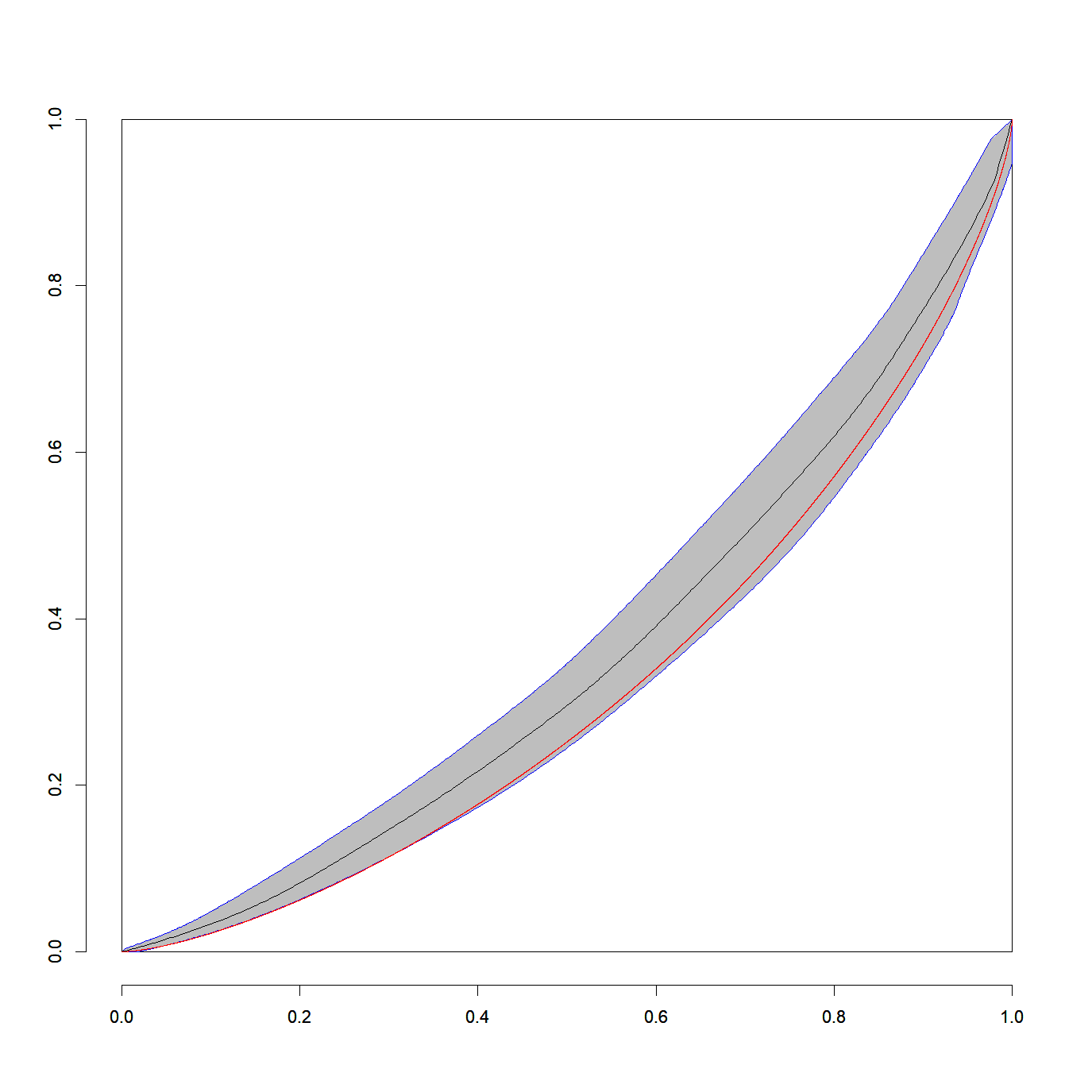}
  \caption{SCB (n=100, $u_\alpha=11.023$).}
  \label{fig:100}
\end{subfigure}%
\begin{subfigure}{.45\textwidth}
  \centering
  \includegraphics[width=\linewidth]{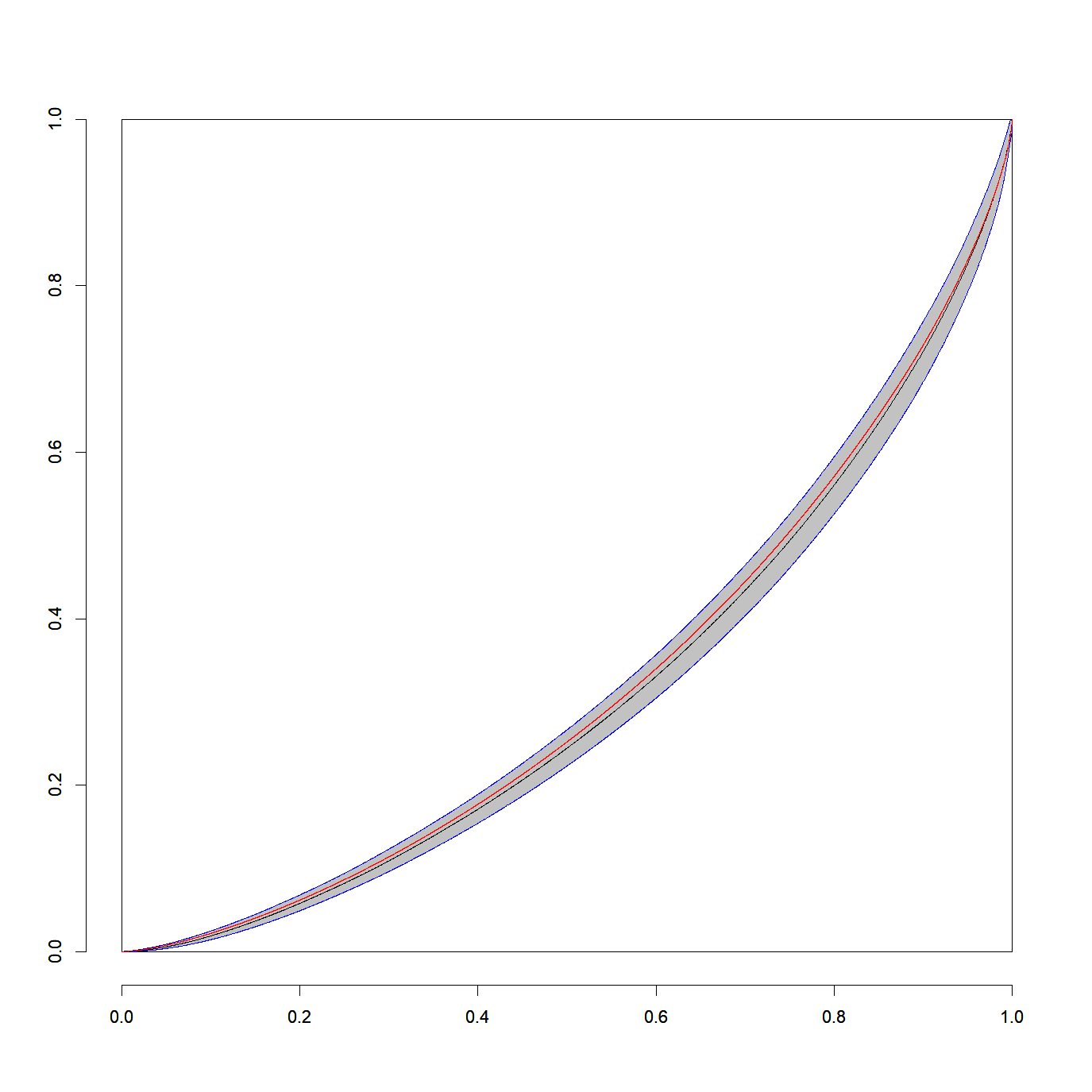}
  \caption{SCB (n=1000, $u_\alpha=14.723$).}
  \label{fig:1000}
\end{subfigure}
\caption{SCBs constructed based on two samples with size n=100 and 1000, respectively, drawn by SRSWOR using the 2015 IT-SILC data at 95\% confidence level.}
\label{fig:1001000}
\end{figure}

Figure~\ref{fig:10p} provides a graphical illustration of the PCB and SCB constructions, as theoretically defined in Expressions~(\ref{eq:exppcb}) and (\ref{eq:expscb}).  These constructions follow the algorithm outlined in Table 1, using the same sample of size 10. The confidence level is 95\%. For $\chi^2_2$, the critical value $c_\alpha$ when $\alpha=0.05$ is 5.991. The adjusted $u_\alpha$ for SCB construction based on the sample is 9.689.
The red curves represent the Lorenz curves at the population level, identical to the one depicted in Figure~\ref{figlb}. The black curves are the estimated Lorenz curves. The blue curves in Figure~\ref{fig:expcb} and \ref{fig:exscb} are the constructed PCB and SCB, respectively. The confidence ellipses are the shaded areas in the grey color, and are the building blocks of PCB and SCB. As seen from the plots, PCB and SCB are the unions of the confidence ellipses. Figure~\ref{fig:10p} is a successful example in that PCB fails to cover entirely the population Lorenz curve, while SCB succeeds. Figure~\ref{fig:1001000} illustrates the SCBs built for the sample Lorenz curves estimated from two samples with different sample sizes, $n=100,1000$. Seen together with Figure~\ref{fig:exscb}, which represents the SCB for a sample of size 10: the larger the sample size is, the narrower the SCB is.

Simulation studies are performed based on sample sizes of $n=10, 100, 200, 500, 1000$. The sample selection is repeated 500 times for each sample size to compute the empirical coverage rates of the PCBs and SCBs for the entire population Lorenz curve. The results are shown in Table~\ref{emp}. At a first glance, the SCB coverage rates seem to be good for small sample sizes, but are bad when sample sizes become large. For example, the SCBs cover below 90\% of the true population Lorenz curve when the sample size is 500 and below 85\% for the sample size of 1000. It is expected that the PCBs will not cover the population Lorenz curve accurately for all sample sizes as they do not account for simultaneous inference in the case of PCBs.

\begin{table}[htb!]
\centering
\begin{threeparttable}
\caption{Empirical coverage rates of the entire population Lorenz curve for a 95\% level of confidence for different sample sizes for PCB (first row) and SCB (second row). }
\begin{tabular}{lrrrrrr}
\toprule
\multicolumn{1}{c}{ } & \multicolumn{5}{c}{Sample size} \\
\cmidrule(l{3pt}r{3pt}){2-6}
  & 10 & 100 & 200 & 500 & 1000 \\
\midrule
PCB & 0.840 & 0.715 & 0.575 & 0.480 & 0.515 \\
SCB & 0.950 & 0.962 & 0.944 & 0.875 & 0.839 \\
\bottomrule
\end{tabular}
\label{emp}
\end{threeparttable}
\end{table}

The undercoverage of the SCBs for large sample sizes is further investigated and understood. It is not the result of a deficiency in the construction method of the SCBs. Instead, the undercoverage problem lies around the endpoints of the estimated Lorenz curve stated in the beginning of Section~\ref{sec:scb}. 
To verify this, we remove 2.5\% of the population Lorenz curve at both endpoints and re-perform the simulation studies. This time, the SCB coverage rates are reasonable for all sample sizes. 

\begin{table}[htb!]
\centering
\begin{threeparttable}
\caption{Empirical coverage rates of the population Lorenz curve skipping 2.5\% of the curve at both endpoints for a 95\% level of confidence}
\begin{tabular}{lrrrrrr}
\toprule
\multicolumn{1}{c}{ } & \multicolumn{5}{c}{Sample size} \\
\cmidrule(l{3pt}r{3pt}){2-6}
  & 10 & 100 & 200 & 500 & 1000 \\
\midrule
PCB & 0.886 & 0.717 & 0.649 & 0.802 & 0.816 \\
SCB & 0.952 & 0.962 & 0.944 & 0.940 & 0.966 \\
\bottomrule
\end{tabular}
\label{edge}
\end{threeparttable}
\end{table}

Nevertheless, imperfections still remain for all sample sizes as shown in Table~\ref{edge} (slight over-coverage for sample sizes of 10, 100 and 1000 and slight undercoverage for sample sizes of 200 and 500). These could be the consequences of several factors. The first layer of the errors might come from the violation of the bivariate normality assumptions imposed on the estimators and the approximation error using the \citet{davies1987hypothesis} method. Secondly, the linearization method for the variance-covariance matrices estimation introduces errors. Thirdly, the implementation of SCB construction based on the algorithm of Table~\ref{alg} results in errors arising from numerical computation. However, despite of these imperfections, the empirical coverage rates of SCBs are sensible and good after removing 5\% of the curve around the endpoints, and are close to 95\% for all sample sizes, which demonstrates the practicality of the proposed method of SCB construction. Since bivariate normality of the point estimators is the only assumption imposed for the SCB construction, the proposed method can be easily extended and applied for building SCBs for errors-in-variables curves of the same or similar kind. It introduces a novel and generic method for SCB construction, applicable as long as point estimators on the curve can be established.

\section{Conclusions \label{sec:con}} 
In this paper, we devise an approach of building SCBs for errors-in-variables curves.
We apply our method to the Lorenz curve by considering it as
an errors-in-variables curve and provide illustration on how to build SCB for it. We determine the form of SCB as the joined confidence ellipses of each point estimator on the curve with individual confidence level adapted for achieving a targeted simultaneous confidence. To the best of our knowledge, it is a novel method of tackling such problem in the research literature.

The paper also includes additional contributions as byproducts. For example, we generalize the linearization method to estimate the covariances and cross-covariances of the estimators.
We also advocate the use of statistical graphs, such
as the Lorenz curve, by providing a practical confidence band to visualize and gain an overall understanding of the data. The methods developed in this paper can be applied to errors-in-variables curves, provided that point estimators for such curves can be established.
Since its introduction, the Lorenz curve has been the focus of numerous studies.
Our research addresses gaps in the literature regarding SCB construction, variance estimation, and the Lorenz curve.

\section*{Acknowledgement}
The authors would like to thank the Swiss National Science Foundation (SNSF) grant 222033 
and 208249
and the Japan Society for the Promotion of Science (JSPS) 
for supporting the research. 
Ziqing Dong acknowledges the constructive discussions with Prof. Yanyuan Ma, Prof. Stefano Peluso and  Prof. Yves Tillé for the research.

\section*{Appendix}
\subsection*{Proofs of Proposition~\ref{lin_Lk1} and Proposition~\ref{lin_Lk2}}
The proofs of Proposition~\ref{lin_Lk1} and Proposition~\ref{lin_Lk2} are a simple application of the quotient rule:
\begin{align*}
\forall i \in S,        \; 
\hat{z}_{k(1	)i} :
&=\frac{\partial \widehat{L}_{k(1)}}{\partial (w_ia_i)}
=\frac{\partial}{\partial (w_ia_i)}\left(\frac{\sum\limits_{j \in U}w_ja_j\1{\scriptstyle[y_j \le y_k]}}{\sum\limits_{j \in U}w_ja_j}\right)\\
&=\frac{\1{\scriptstyle[y_i \le y_k]}\bigg(\sum\limits_{j \in U}w_ja_j\bigg)-\bigg(\sum\limits_{j \in U}w_ja_j\1{\scriptstyle[y_j \le y_k]}\bigg)}{\bigg(\sum\limits_{j \in U}w_ja_j\bigg)^2}\\
&=\frac{\1\scriptstyle[y_i \leq y_k]}{\bigg(\sum\limits_{j \in U}w_ja_j\bigg)}
-\frac{1}{\bigg(\sum\limits_{j \in U}w_ja_j\bigg)}\widehat{L}_{k(1)}\\
&=\frac{\1\scriptstyle[y_i \leq y_k]}{\widehat{N}}
-\frac{1}{\widehat{N}}\widehat{L}_{k(1)}.
\end{align*}
Similarly, we have 
\begin{align*}
\forall i \in S,        \; 
\hat{z}_{k(2	)i} :
&=\frac{\partial \widehat{L}_{k(2)}}{\partial (w_ia_i)}
=\frac{\partial}{\partial (w_ia_i)}\left(\frac{\sum\limits_{j \in U}w_jy_ja_j\1{\scriptstyle[y_j \leq y_k]}}{\sum\limits_{j \in U}w_jy_ja_j}\right)
\\
&=\frac{\bigg(y_i\1{\scriptstyle[y_i \le y_k]\bigg)}\bigg(\sum\limits_{j \in U}w_jy_ja_j\bigg)-\bigg(\sum\limits_{j \in U}w_jy_ja_j\1{\scriptstyle[y_j \le y_k]}\bigg)y_i}{\bigg(\sum\limits_{j \in U}w_jy_ja_j\bigg)^2}\\
&=\frac{y_i\1\scriptstyle[y_i \leq y_k]}{\bigg(\sum\limits_{j \in U}w_jy_ja_j\bigg)}
-\frac{y_i}{\bigg(\sum\limits_{j \in U}w_jy_ja_j\bigg)}\widehat{L}_{k(2)}\\
&=\frac{y_i\1{\scriptstyle[y_i \leq y_k]}}{\widehat{Y}}
-\frac{y_i}{\widehat{Y}}\widehat{L}_{k(2)}.
\end{align*}
Obviously, 
$\forall i \in U,$ we have
$$z_{k(1)i} :
=\hat{z}_{k(1)i}\Bigg|_{\substack{\begin{subarray}{l}\mathlarger{\mathbf{w \hspace{-1mm} \circ \hspace{-1mm} a}}\\=
\mathlarger{\bold 1}\end{subarray}}}
=\frac{\1\scriptstyle[y_i \leq y_k]}{N}
-\frac{1}{N}L_{k(1)},$$ 
and 
$$z_{k(2)i} :
=\hat{z}_{k(2)i}\Bigg|_{\substack{\begin{subarray}{l}\mathlarger{\mathbf{w \hspace{-1mm} \circ \hspace{-1mm} a}}\\=
\mathlarger{\bold 1}\end{subarray}}}
=\frac{y_i\1{\scriptstyle[y_i \leq y_k]}}{Y}
-\frac{y_i}{Y} L_{k(2)}.$$

\bibliographystyle{Chicago}
\bibliography{bib}
\end{document}